\documentclass[conference]{IEEEtran}
\IEEEoverridecommandlockouts
\usepackage{etoolbox}
\usepackage{tikz}

\usepackage{listings}

\usepackage{graphicx} 
\usepackage{enumitem}
\usepackage{circledsteps}
\usepackage{subcaption}
\usepackage{multirow}
\usepackage{hhline}
\usepackage{lipsum}
\usepackage{rotating}
\usepackage{algorithm}
\usepackage{algpseudocode}
\usepackage{soul}
\usepackage{cite}
\usepackage{amsmath,amssymb,amsfonts}
\usepackage{graphicx}
\usepackage{textcomp}
\usepackage{xcolor}
\usepackage{tcolorbox}
\usepackage{booktabs}
\newbool{showComments}
\booltrue{showComments}
\ifbool{showComments}{%
\newcommand{\rob}[1]{\textcolor{blue}{\textbf{RM:} #1}}
\newcommand{\sy}[1]{\textcolor{orange}{\textbf{SJ:} #1}}
\newcommand{\deleteifnospace}[1]{{\textcolor{red}{#1}}}

}{
\newcommand{\rob}[1]{}
\newcommand{\sy}[1]{}
\newcommand{\deleteifnospace}[1]{}

}
\usepackage[colorlinks,urlcolor=blue,linkcolor=blue,citecolor=blue]{hyperref}
\usepackage{comment}
\expandafter\def\expandafter\UrlBreaks\expandafter{\UrlBreaks\do\/\do\*\do\-\do\~\do\'\do\"\do\-}

\begin{document}

\title{Smaller, Smarter, Closer:\\ The Edge of Collaborative Generative AI}

\author{\IEEEauthorblockN{Roberto Morabito}
\IEEEauthorblockA{\textit{EURECOM} \\
Biot, France \\
roberto.morabito@eurecom.fr}
\and
\IEEEauthorblockN{SiYoung Jang}
\IEEEauthorblockA{\textit{Nokia Bell Labs} \\
Cambridge, United Kingdom \\
siyoung.jang@nokia-bell-labs.com}
}

\maketitle

\begingroup
\renewcommand\thefootnote{}\footnotetext{\noindent\footnotesize This paper has been accepted for publication in \textit{IEEE Internet Computing}. Upon publication, the copyright will be transferred to IEEE.}%
\addtocounter{footnote}{-1}
\endgroup

\begin{abstract}
The rapid adoption of generative AI (GenAI), particularly Large Language Models (LLMs), has exposed critical limitations of cloud-centric deployments, including latency, cost, and privacy concerns. Meanwhile, Small Language Models (SLMs) are emerging as viable alternatives for resource-constrained edge environments, though they often lack the capabilities of their larger counterparts. This article explores the potential of collaborative inference systems that leverage both edge and cloud resources to address these challenges. 
By presenting distinct cooperation strategies alongside practical design principles and experimental insights, we offer actionable guidance for deploying GenAI across the computing continuum.
Ultimately, this work underscores the great potential of \textit{edge-first} approaches in realizing the promise of GenAI in diverse, real-world applications.

\end{abstract}

\maketitle


\section{Introduction}
It is no longer necessary to elaborate extensively on the transformative impact of generative AI (GenAI) models, particularly Large Language Models (LLMs), across various sectors of society. From healthcare to education, entertainment to software development and IoT \cite{10637270}, it is evident that nearly every application domain is ready (or already is) to be influenced by these technologies. LLMs like GPT-4, powered by transformer architectures with billions of parameters, excel in diverse NLP tasks (e.g., summarization, translation, query answering) and high-level reasoning. In this paper, we often refer to language models as the most advanced and widely adopted category of GenAI. However, the considerations and strategies discussed are equally relevant to other types of GenAI models, such as those used for image generation or multimodal applications, as their characteristics and challenges often overlap with those of language models. For example, models such as ChatGPT-4o achieve top benchmarks in \textit{general knowledge} (MMLU), \textit{math reasoning} (GSM8K), and \textit{code generation} (HumanEval) while integrating multimodal capabilities \cite{hurst2024gpt}, redefining machine potential in modern digital ecosystems. These capabilities come with high computational and storage demands, typically managed via cloud hosting. 

While the transformative potential of GenAI is undeniable, from the perspective of distributed AI and systems, significant \textit{pre-GenAI era} efforts aimed to shift AI execution closer to the edge. The motivations that drove these efforts—reducing latency, enhancing reliability, and improving user experiences—remain equally relevant in the context of GenAI. However, with LLMs, for example, additional challenges emerge. A persistent issue with cloud-based LLMs is the high cost of provisioning these services \cite{luccioni2023estimating}. Beyond costs, there are growing concerns regarding security and privacy. For instance, content submitted to popular LLM platforms—such as user prompts, model responses, and other data—is often used to improve model performance. This occurs routinely, as services like ChatGPT have become as ubiquitous as web search engines, with little thought given to the implications of such data sharing. Moreover, the centralized nature of cloud-based LLMs introduces risks of service disruption. For example, a four-hour outage of OpenAI’s services caused significant disruptions for applications heavily reliant on its APIs \cite{openai_incident_2024}, underscoring the reliability challenges of centralized systems. These issues are further compounded by the impact of GenAI on network traffic, as highlighted in \cite{ericsson2024GenAI}. Increased use of GenAI-driven video assistants and immersive interactions is expected to drive significant growth in both uplink and downlink traffic, placing additional strain on network infrastructure. On the other hand, GenAI's compression capabilities are likely to find utility in closed ecosystem applications, including edge-based deployments, where they could play a key role in reducing traffic demands by processing and semantically compressing data closer to users, thereby alleviating strain on network infrastructure.
These concrete issues from the perspective of \textit{GenAI over the Internet} are driving the push toward edge-based solutions.

\begin{figure}[ht]
    \centering
    \includegraphics[width=.8\columnwidth]{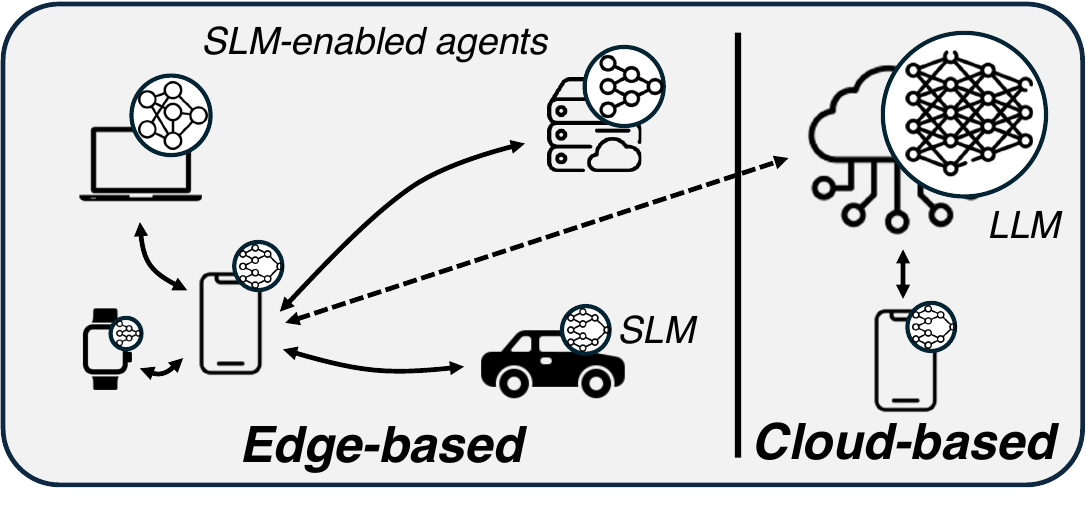}
    \caption{Interaction between edge-based and cloud-based SLM-enabled agents.}
    \label{fig:overview}
\end{figure}

As the reliance on edge-based solutions grows, Small Language Models (SLMs) are emerging as a key enabler for addressing these challenges through lightweight, \textit{agentic} systems designed for collaborative and distributed AI, as highlighted in \cite{meuser2024revisiting}. Adopting SLMs at the edge, therefore, should not be viewed solely as an engineering or data science endeavor to shrink models while maintaining quality. Instead, from a distributed AI standpoint, this shift represents an opportunity to address these challenges holistically.

While SLMs are the way forward to efficiently bring GenAI to resource-constrained environments, they do not fully resolve the challenges associated with deploying AI at the edge. The limitations of SLMs, particularly in delivering the qualitative and quantitative performance of their larger counterparts, make it unlikely that the cloud can be completely replaced \cite{laskaridis2024future}. Instead, we argue for a complementary relationship between the edge and the cloud (as exemplified in Figure \ref{fig:overview}), where SLM-enabled agents deployed at the edge collaborate not only with centralized infrastructures, but also among themselves. This approach prioritizes the edge as the primary execution environment, relying on the cloud only when QoE from the end user's perspective and QoS from the system’s capabilities cannot be fully satisfied at the edge.
Furthermore, this collaboration introduces additional dimensions compared to pre-GenAI applications, where edge cooperation was primarily driven by \textbf{\textit{computation}} needs. In fact, thanks to the versatility and multi-purpose nature of these models, we envision a broader scope for cooperation, encompassing not just computation but also \textbf{\textit{data}} and \textbf{\textit{knowledge}} sharing across heterogeneous systems.

Building on these considerations, this article explores the potential of \textit{collaborative GenAI at the edge}, offering practical insights into designing such systems.
From abstracting how to enable dynamic task delegation across heterogeneous devices to introducing the scalability of distributed LLM inference, our aim is to provide a roadmap for deploying collaborative intelligence across the computing continuum.

\section{From Cloud to Edge: Motivation}
Recent trends reveal a notable shift toward reducing the size of LLMs, with a growing emphasis on compactness and efficiency. This shift is motivated by factors spanning both the capabilities of smaller models themselves and the challenges posed by large models provisioning from a network perspective. In this section, we explore these motivations and their implications for the future of GenAI. 

 \begin{figure}[!t]
    \centering
    \includegraphics[width=\columnwidth]{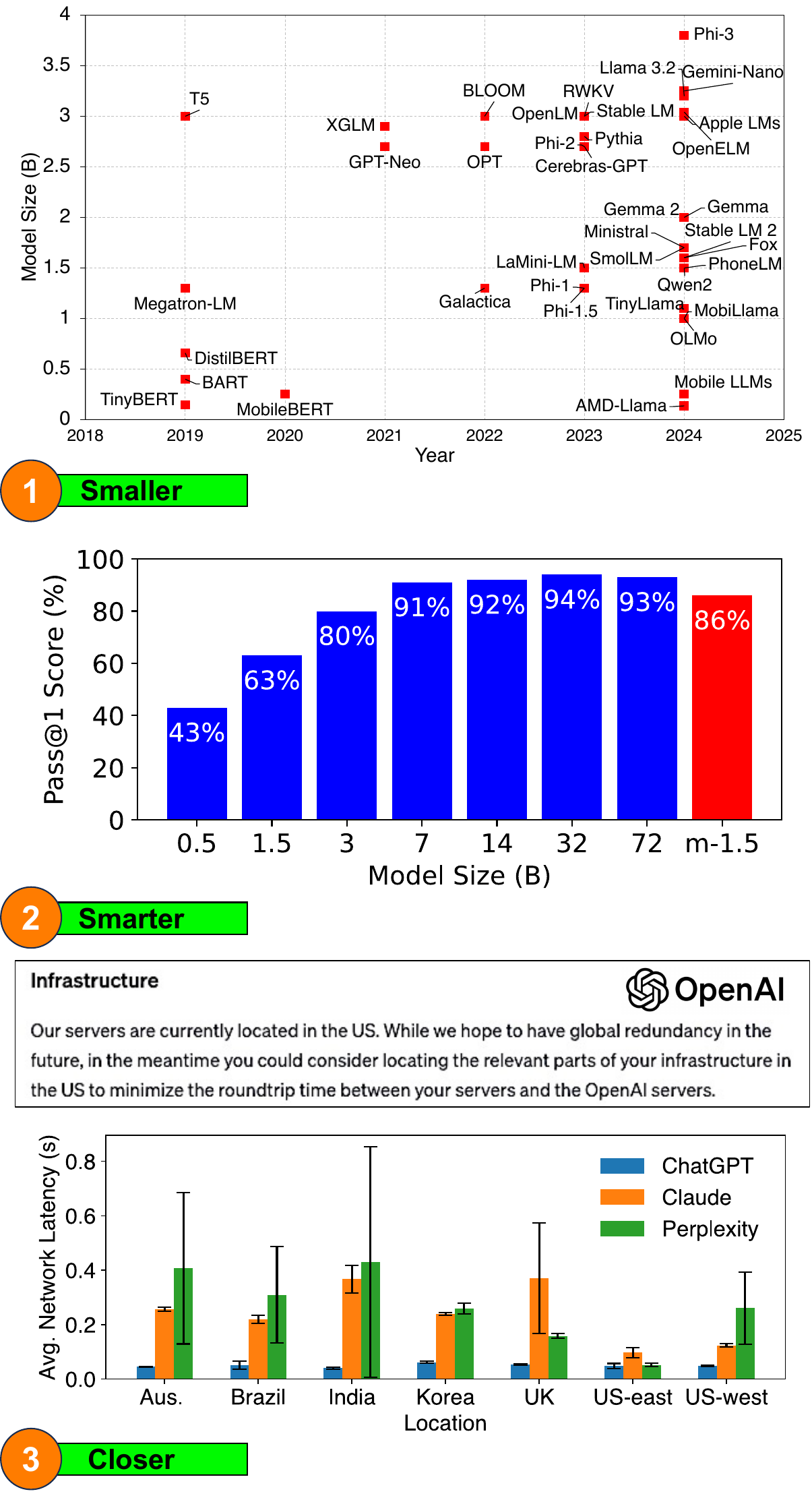}
    \caption{\textit{Smaller, Smarter, Closer}: (1) Progression towards smaller language models with comparable or improved efficiency (top); (2) Improved accuracy of smaller models as measured by Pass@1 scores (middle); and (3) Infrastructure-related latency comparisons across global locations for major AI systems, highlighting geographical disparities in access speed (bottom).}
    \label{fig:sss}
\end{figure}

 \begin{figure*}[ht]
    \centering
    \includegraphics[width=\textwidth]{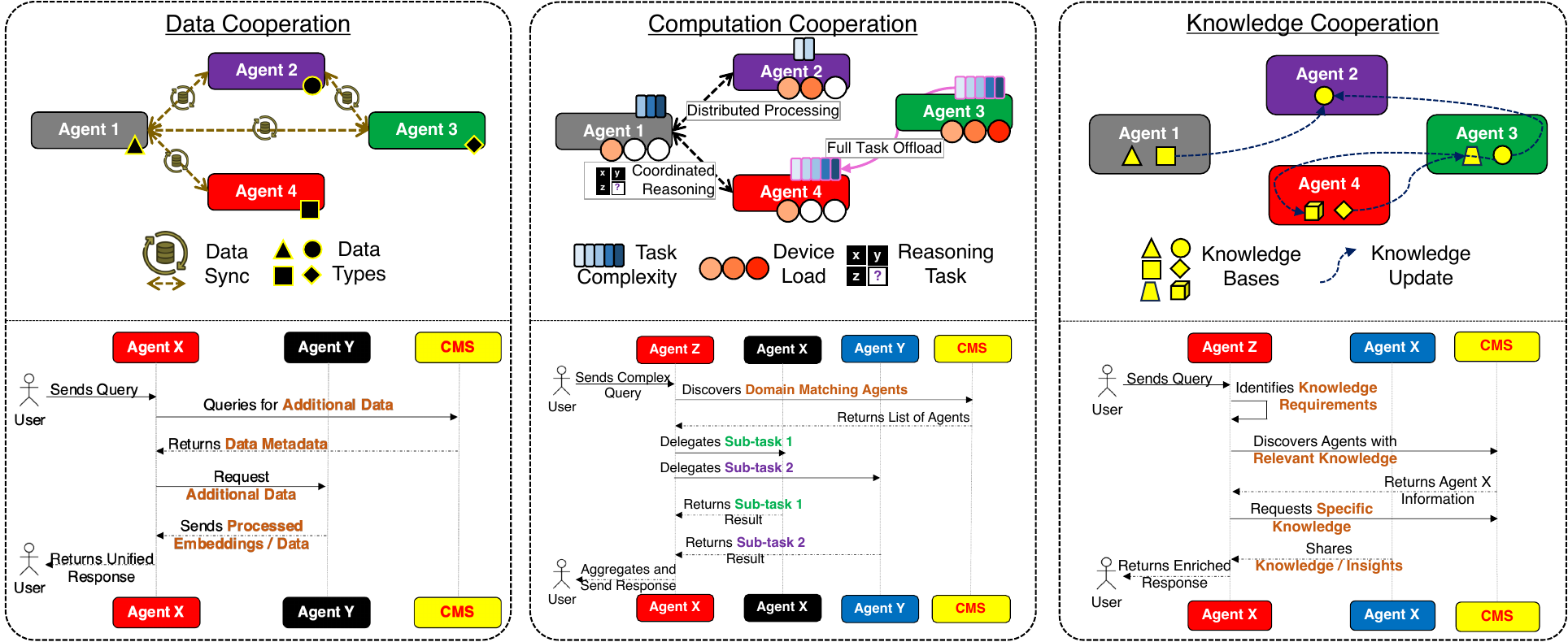}
    \caption{Task-oriented cooperation types: Data, Computation, and Knowledge cooperation strategies for multi-agent collaboration. The Capability Metadata Store (CMS), introduced later in the paper, is depicted here as a key enabler for coordination and metadata-driven interactions among agents.}
    \label{fig:collaboration}
\end{figure*}

\subsection{Smaller} In response to the high cost \cite{luccioni2023estimating} of cloud-based LLM services, advances in model compression techniques—such as quantization \cite{lin2024awq-quantization}, pruning \cite{ma2023llmpruner}, and knowledge distillation \cite{gu2024llm-kd}—are making edge AI increasingly viable for deploying SLMs on mobile devices \cite{laskaridis2024melting} and at the edge.
The upper part of Figure~\ref{fig:sss}, \textit{'Smaller'}, depicts the recent surge in sub-4B language models developed over the past few years. These models vary in size, with the smallest, Qwen2.5, featuring 0.5B parameters and a storage footprint of just 398 megabytes. Notably, these models are also becoming multi-modal, with capabilities extending beyond text processing to include image-to-text capabilities (Llama3.2-vision). This current shift reflects a broader industry trend toward developing \textit{smaller} and versatile models that can handle a range of input types, enabling them to support complex application that require integrating information from multiple data sources–capability especially necessary at the network edge. 
\\
\subsection{Smarter} The challenge with SLMs lies in their reduced computational capability resulting from quantization and compression. Drastically reducing a general-purpose model leads to significant performance degradation.
Blue bar in Figure~\ref{fig:sss}, \textit{'Smarter'}, depicts Pass@1 accuracy (percentage of queries answered correctly on the first try) of OpenAI/GSM8K math test accuracy drop as the number of language model parameters drop. A language model pre-trained on a specific domain may outperform a general-purpose model when handling domain-specific queries. Instead of an all-purpose model, domain-specific models like Code Llama (code generation), DeepSeekMath (mathematical reasoning), and BioMistral (medical Q\&A), each with 7B parameters, achieve greater accuracy within their specialized fields. 
Red bar in Figure~\ref{fig:sss} middle, demonstrate that despite having only 1.5 billion parameters, the math domain specific model \textit{m-1.5} achieves accuracy comparable to that of 7B general-purpose variant while requiring just 19.8\% of the memory. This demonstrates that while SLMs may have fewer parameters and reduced general-purpose capabilities, their domain-specific optimization allows them to achieve exceptional accuracy and efficiency, highlighting the potential for SLMs to become \textit{smarter} and more reliable within targeted domain.
\\
\subsection{Closer}
Long latency is a common issue in cloud-based LLM services, as data must be transmitted over the network, processed remotely, and then returned, introducing delays that impact real-time applications. Figure~\ref{fig:sss}, \textit{'Closer'}, shows network latency measurement of various cloud-based LLM services. 
Popular LLM services, such as those provided by OpenAI, are hosted in centralized server locations (e.g., the US), which can impact the QoE for users situated far from these hosting regions due to increased latency and reliability issues, with network latency ranging from around tens of milliseconds to hundreds of milliseconds depending on the service provider and user location. Content Delivery Network (CDN) systems can help to alleviate long latency delays by efficient network tunneling and management. However, CDNs cannot fully resolve the issue since the core processing of LLMs remains on remote infrastructure. This inherent limitation further drives research into moving LLM services \textit{closer} to the edge, enabling localized inference to significantly reduce networking latency and enhancing QoE for users. 
\\
\section{Edge-Centric Collaboration}
\subsection{Collaboration Approaches}
Advancing collaborative GenAI inference at the edge requires a structured understanding of the diverse strategies enabling agents to cooperate and process tasks efficiently. These strategies address critical aspects of performance, scalability, and adaptability, particularly in resource-constrained environments. As shown in Figure \ref{fig:collaboration}, task-oriented cooperation is categorized into three key types: \textit{Data}, \textit{Computation}, and \textit{Knowledge}. Each category addresses distinct operational challenges: Data Cooperation ensures synchronized observations, Computation Cooperation optimizes resource usage through task distribution, and Knowledge Cooperation enables dynamic access to domain-specific expertise. These approaches collectively form the foundation of scalable and efficient edge-based GenAI systems. Each cooperation type is illustrated in Figure ~\ref{fig:collaboration}, with a high-level conceptual overview at the top and an example showcasing specific interactions between agents below.
\\
\textbf{Data cooperation} ensures agents maintain consistent and up-to-date knowledge through the real-time exchange of raw or processed data. A key enabler is the exchange of embeddings, compact representations that encode domain-specific information \cite{yao2024velo}. By sharing embeddings with metadata (e.g., domain, type, dimensionality), agents can reuse preprocessed data for tasks like classification or summarization, reducing computational overhead and enhancing scalability.

 \begin{figure*}[ht]
    \centering
    \includegraphics[width=\textwidth]{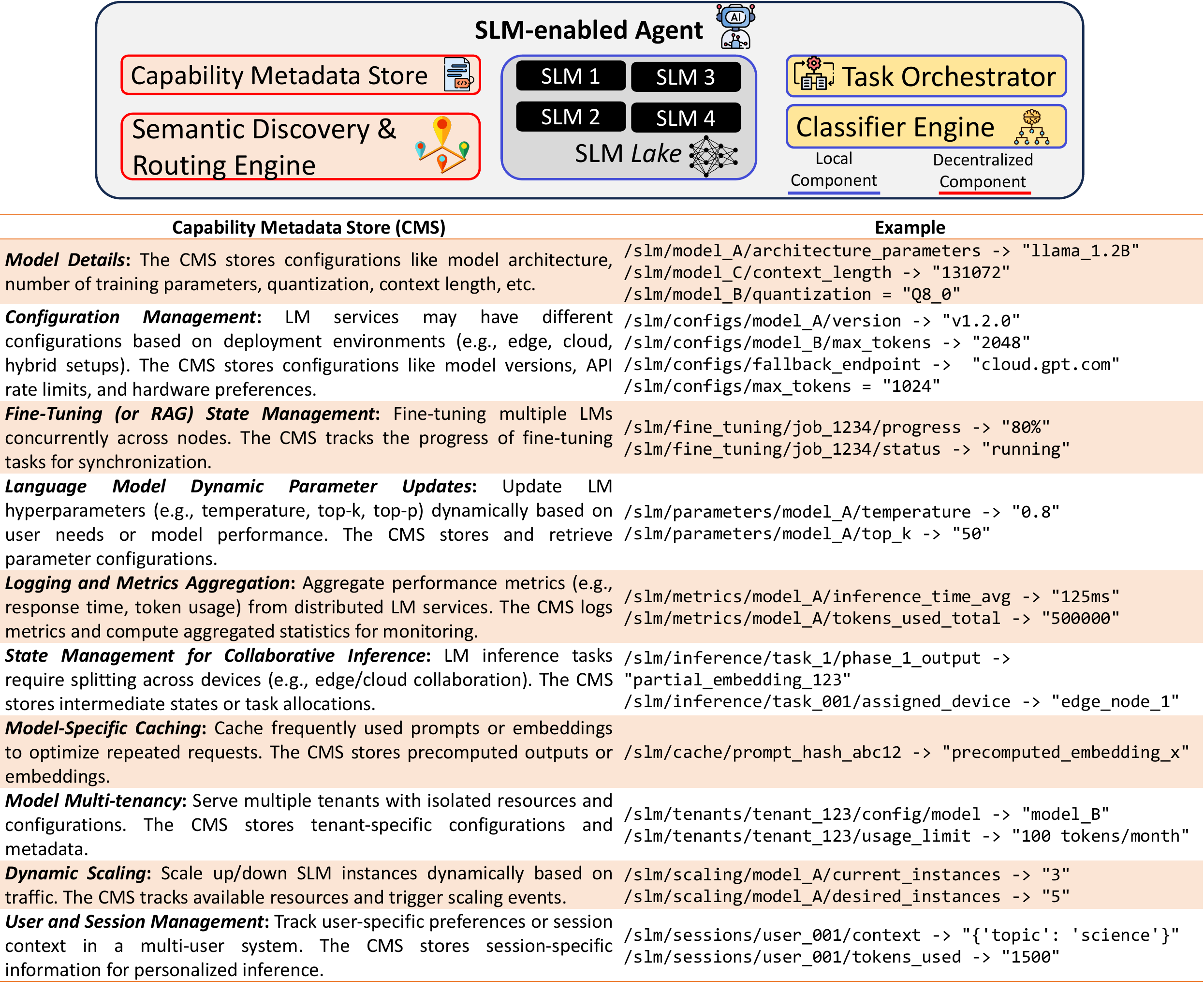}
    \caption{On top, our conceptual framework for enabling collaboration among distributed SLM-enabled agents. On the bottom, the table illustrates examples of CMS entries for coordinating data, computation, and knowledge cooperation across agents.}
    \label{fig:stack_cms}
\end{figure*}
This approach is particularly valuable in situational awareness scenarios, such as environmental monitoring and real-time analytics. For example, an air quality monitoring agent can generate embeddings enriched with metadata (e.g., urban vs. industrial), which are shared with peers for trend detection or anomaly analysis. This synchronization enables agents to construct a unified environmental representation, improving responsiveness while minimizing redundant processing.
\\
\textbf{Computation cooperation} involves dynamically allocating and distributing computational tasks to balance workloads and optimize resource utilization. Unlike data cooperation, which focuses on synchronization, this approach considers both agents’ computational capabilities and the complexity of the request \cite{zhang2024edgeshard}. By matching tasks with the most appropriate resources, agents collaborate effectively. For instance, a complex query can be decomposed into subtasks, with lightweight tasks assigned to lower-capability agents and resource-intensive operations handled by specialized nodes. The results are then aggregated, showcasing how distributed inference pipelines leverage both agent expertise and computational power to handle demanding workloads.
\\
\textbf{Knowledge cooperation} addresses knowledge gaps by enabling agents to access domain-specific expertise or general-purpose reasoning capabilities \cite{das2023enabling}. Unlike data cooperation, which focuses on synchronization, and computation cooperation, which emphasizes execution, knowledge cooperation facilitates the dynamic exchange of higher-order insights for deeper contextual understanding. For instance, an agent processing a math or code query may detect a knowledge gap and query a peer with domain-specific expertise. Metadata ensures the retrieved knowledge is accurate, contextually aligned, and relevant, allowing it to be seamlessly integrated into operations.
Beyond addressing immediate gaps, this approach supports long-term adaptability. Through iterative exchanges, agents refine their shared knowledge base, enhancing their capabilities and decision-making over time. This evolving collaboration is particularly valuable for reasoning-intensive tasks in dynamic and complex environments, where a unified and adaptive approach is essential.

\subsection{Requirements}
Seamless collaboration among distributed agents requires understanding the diverse dimensions of such systems, which operate in heterogeneous environments with varying computational capacities, network latencies, and data domains. Instead of proposing a one-size-fits-all architecture—neither feasible nor suitable—this effort emphasizes foundational principles and mechanisms for achieving flexible, adaptable, and scalable collaboration.

Building on the cooperation strategies discussed earlier, successful systems must prioritize \textbf{flexibility in capability management}. This involves enabling agents to dynamically discover and interpret peers' capabilities, such as sharing data or leveraging domain-specific expertise. Metadata plays a pivotal role, guiding decisions about the relevance of an agent’s resources for specific tasks. For example, an agent specializing in healthcare could tag outputs with metadata describing their domain and type, allowing others to reuse these resources efficiently.

Another requirement is \textbf{adaptability to task complexity}, given the variety of SLM applications. Tasks can range from simple fact retrieval to complex reasoning workflows. Systems must support task decomposition, where a query is split into subtasks handled collaboratively by agents. For instance, one agent might extract context while another performs detailed inferences, ensuring collaboration meets latency, accuracy, or resource efficiency priorities.

Collaborative systems must also address \textbf{knowledge gaps}, particularly in scenarios where agents encounter queries beyond their expertise. Agents should detect insufficient knowledge, query peers with relevant expertise, and seamlessly integrate external insights into their inference processes. For example, an agent handling a marine biology query might dynamically retrieve domain-specific knowledge from a peer, ensuring contextually aligned and accurate responses.

Finally, practical implementations of distributed GenAI inference must \textbf{balance trade-offs} between competing priorities. Systems must optimize performance while avoiding resource overburden and manage decentralization's scalability benefits alongside its coordination overhead.

\subsection{Elements Supporting Collaborative SLM Inference}

To address the outlined requirements, we identify key elements essential for enabling collaboration among distributed SLM-enabled agents. The upper part of Figure \ref{fig:stack_cms} illustrates the conceptual framework of these components. While we do not aim to propose a fully-fledged architecture, this figure offers an overview of key building blocks that align with our view

The \textit{CMS} acts as a decentralized repository for storing and retrieving metadata on agent capabilities, including model configurations, system nodes, and resources. As shown on the bottom of Figure \ref{fig:stack_cms}, it contains entries such as model architecture, parameters, quantization settings, and system metrics like task progress, node health, and load status.  Information sharing among devices is performed through gossip-like protocols \cite{zhang2020distributed}, favoring broad and adaptive metadata dissemination across agents. 
Overall, all these different entries provide a structured way to enable dynamic service discovery and metadata-driven decision-making, as discussed earlier.

\textit{Semantic Discovery and Routing} capabilities are also essential. Adopting a simple, semantically structured notations akin to DNS-style naming conventions (e.g., \texttt{data.gpt/domain/processed\_embeddings} for data, \texttt{computation.gpt/task\_distribution} for computations, or \texttt{knowledge.gpt/arithmetic} for domain-specific knowledge), agents can quickly identify and interact with relevant services. This approach simplifies coordination across heterogeneous systems and facilitates seamless interoperability.

The \textit{Task Orchestrator} complements this by managing task decomposition and dynamic allocation of subtasks across agents. It leverages information from the CMS and classifier outputs to adaptively balance workloads, ensuring that computational resources are used efficiently. The \textit{Classifier Engine} \cite{yadav2024pag-llm}, in turn, runs specialized classifiers to assess task complexity and relevance, gap detection, and capability matching, allowing the orchestration process to align with the adaptability and scalability principles highlighted in the requirements section.

These components can collectively support the execution of the collaboration strategies discussed earlier—data, computation, and knowledge cooperation—by providing the necessary infrastructure for agents to interact cohesively. They leverage structured metadata, semantically simple discovery notations, and adaptive task orchestration, ensuring consistent coordination across heterogeneous environments.

Once again, while this framework is not intended to be exhaustive, it can provide a baseline for designing collaborative systems that fulfill the requirements outlined previously.

\subsection{Security and Privacy Considerations}

Before moving forward to practical application scenarios, it is important to briefly acknowledge security and privacy considerations that arise in collaborative edge inference scenarios. While these systems offer benefits in latency reduction and resource efficiency, they also introduce potential security vulnerabilities. Sensitive information could be leaked during inter-agent communication or task delegation, and adversaries could target intermediate inference states. Addressing these challenges requires incorporating privacy-preserving mechanisms, such as encrypted data exchanges, secure multi-party computation, and fine-grained access controls. Further discussions on attacks and defense mechanisms in collaborative inference systems can be found in \cite{he2020attacking}.

\section{Application Scenarios}
Next, we analyze practical use case scenarios for collaborative SLMs-enabled agents at the edge, demonstrating how the previously discussed cooperation strategies and components can be applied.

\subsection{Mobile \& Wearable Healthcare. } \label{subsec:healthcare}

\begin{figure}[!t]
    \centering
    \includegraphics[width=.85\columnwidth]{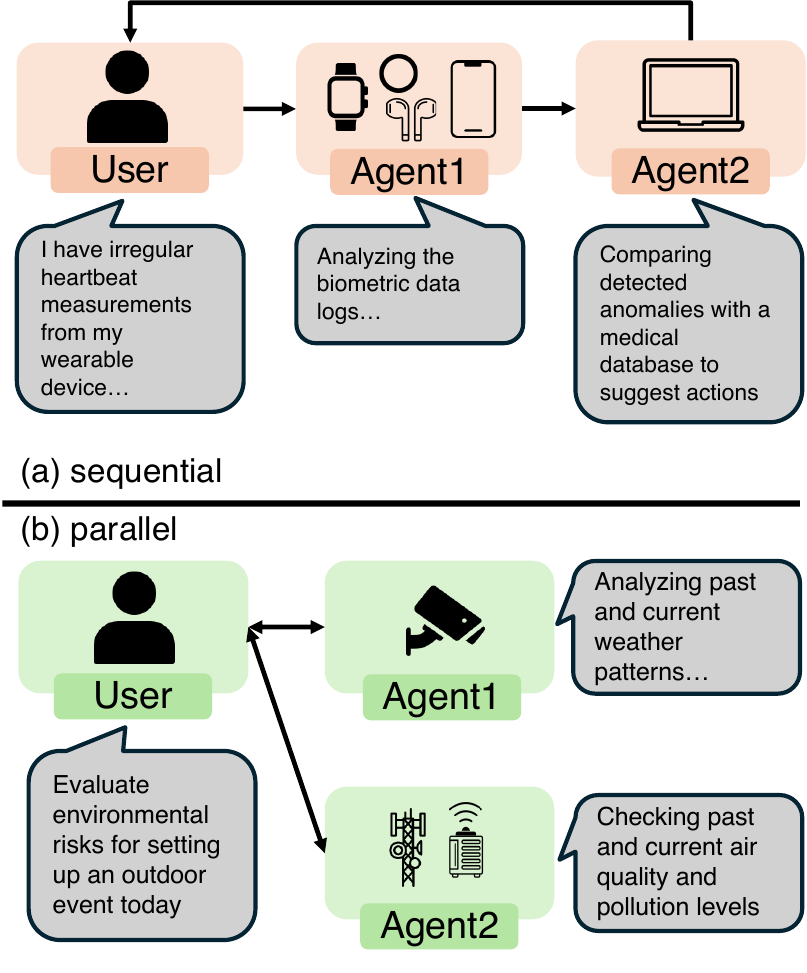}
    \caption{A collaborative language model inference scenario illustrating (a) sequential collaboration and (b) parallel collaboration.}
    \label{fig:collaboration_scenario}
\end{figure}

Wearable AI enables real-time health monitoring, integrating multi-modal data for personalized care. The advent of multi-modal GenAI models, including SLMs, further enhances these capabilities by enabling devices to integrate multiple functions within a single model. This allows for context-aware analysis, collaborative decision-making, and more efficient use of on-device resources.

This is particularly valuable as the number of AI-enabled wearables on our bodies increases~\cite{gong2023collaborative}, along with the variety of available sensors. These devices now provide a much broader range of sensory data, including heart rate, respiration rate, blood oxygen level, activity levels, sleep patterns, and environmental conditions. Each wearable device can process such data locally using lightweight SLM models, transforming raw sensor inputs into actionable insights. 
This capability reduces reliance on centralized processing systems, enhancing the overall efficiency and scalability of health monitoring solutions from a technical perspective. From an application perspective, it enables devices to provide deeper insights into a person's health status by integrating and analyzing diverse data streams collectively gathered from multiple devices.

As depicted in Figure~\ref{fig:collaboration_scenario} (a), in a scenario where a user queries their symptoms, multiple on-body computing devices can sequentially and collaboratively perform SLM inference by leveraging a combination of data and knowledge cooperation. These agents analyze real-time vitals, share insights to create a unified understanding of the user's health status, and dynamically query relevant knowledge bases when needed. This enables the system to proactively provide personalized advice to the wearer while auto-generating detailed reports for healthcare providers, all without requiring the user's intervention.

\subsection{Collaborative Urban Intelligence}
In public spaces, ML model-equipped agents can seamlessly integrate into the smart city ecosystem, embedding themselves within public infrastructures such as smart cameras, environmental sensors, traffic lights, and roadside units. Each of these systems is equipped with processing power and specialized onboard sensors—for instance, visual and audio sensors on public smart cameras or environmental sensors on roadside units—enabling them to function intelligently within urban infrastructure. 
Imagine a scenario in the context of increasingly integrated smart cities where a user's device cannot accurately address a query using only its onboard data, model, and information. The device can enhance its understanding and decision-making capabilities, by collaborating with nearby IoT/Edge devices. 

As depicted in Figure~\ref{fig:collaboration_scenario} (b), a user's device can query their device with a question such as \textit{Evaluate environmental risks for setting up an outdoor event}. In response, the user's device decompose the root query and start collaborating with nearby IoT agents, leveraging both computation cooperation and data cooperation to provide a comprehensive answer. Each of these queries can then be concurrently addressed with high accuracy by specific IoT devices. 
For example, a smart camera can leverage its visual language model and its data to report recent observation while also conducting an online search for records in the area. Simultaneously, an environment monitoring identifies recent pollution levels. The aggregated insights allow the user's device to form a holistic understanding of the environment, enabling better-informed decisions about urban safety and quality of life. 

\begin{figure}[!t]
    \centering
    \includegraphics[width=1\columnwidth]{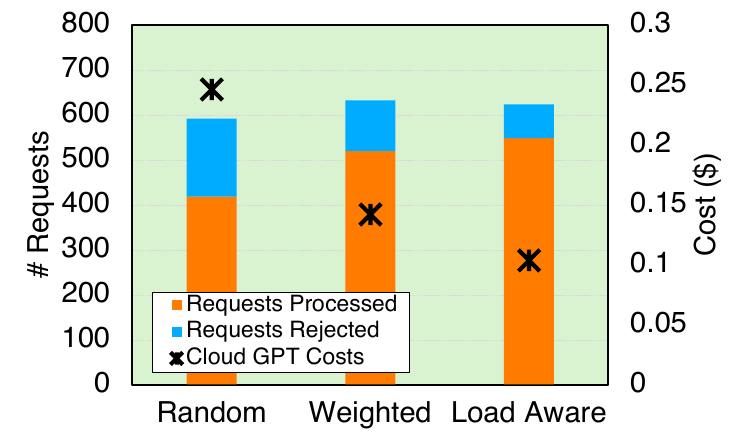}
    \caption{Impact of request allocation methods on the number of requests processed, rejected, and cloud redirection costs in collaborative edge-cloud systems.}
    \label{fig:cloud_redirections}
\end{figure}

\section{Balancing Edge Constraints And Cloud Dependencies} While cooperation strategies enable efficient edge deployments, many edge systems face inherent computational limitations \cite{zhou2019edge}. As demand rises, edge devices can quickly reach capacity limits, making cloud-based GenAI a necessary complement, particularly when edge resources saturate. However, reliance on cloud services introduces non-negligible costs, emphasizing the need for efficient scheduling and orchestration to minimize cloud redirections \cite{zhangefficient}.

To evaluate this, we deployed a system consistent of eight off-the-shelf edge devices, including NVIDIA Jetson AGX Orin and Jetson Orin Nano\footnote{https://www.nvidia.com/en-us/autonomous-machines/embedded-systems/}, featuring built-in GPUs. Each device had predefined token generation (where a \textit{token} represents a unit of text processed during inference) and concurrency limits, reflecting constraints derived from benchmarking GenAI workloads \cite{laskaridis2024melting}.

The system configuration was tailored to replicate real-world conditions. Over a one-hour evaluation, the system was subjected to bursty workloads, with periodic surges in user requests. Token demands for each request were selected from typical sizes (e.g., 50, 100, 200, 300, and 500 tokens), weighted toward smaller requests to reflect common usage patterns. Cloud offloading costs were based on OpenAI API GPT-4 pricing\footnote{https://openai.com/api/pricing/}, accounting for input and output tokens.

A critical challenge in such deployments is optimally assigning requests to nodes. We evaluated three request-allocation methods: \textit{Random}, which distributes requests uniformly across nodes regardless of capacity; \textit{Weighted}, which prioritizes higher-capability nodes; and \textit{Load-Aware}, which dynamically directs requests to the least utilized nodes, balancing the workload and mitigating bottlenecks.

As shown in Figure~\ref{fig:cloud_redirections}, scheduling strategies significantly impact system performance and cloud redirections. The Random method led to uneven workload distribution, increasing task rejections and cloud reliance. In contrast, Weighted allocation reduced cloud offloading by prioritizing higher-capability nodes, while Load-Aware scheduling further improved efficiency by dynamically balancing the workload, minimizing rejected tasks, and lowering cloud costs.

These results are not intended to lead to generalizations, as edge deployments can vary widely in their configurations, the GenAI applications they provision, and the degree of heterogeneity in their infrastructure. However, they underscore the significance of even a simple change in scheduling strategy, which in this case halved cloud costs. This reduction becomes even more impactful when considering the hyperscale nature of the use cases discussed earlier, where millions of requests may be processed daily.

These findings open the door to numerous research opportunities. Advancing scheduling algorithms that dynamically adapt to GenAI workload variations and edge heterogeneity will be paramount in shaping the next generation of edge intelligence networks. Additionally, investigating their scalability across diverse deployment scenarios and multi-modal applications could further enhance real-world applicability. Exploring the feasibility of interoperability among heterogeneous SLMs, devices, platforms will also be critical in ensuring seamless integration across the ecosystem. Finally, we envision that incorporating \textit{incentive mechanisms} could foster deeper collaboration between edge and cloud resources, ensuring that stakeholders are fairly compensated for their contributions, thereby strengthening the overall ecosystem of collaborative GenAI inference.

\section{Conclusion}
Edge computing will play a key role in addressing the limitations of cloud-centric GenAI, favoring low-latency, cost-effective, and privacy-preserving solutions. In this article, we explored collaboration strategies and practical design principles for deploying GenAI across the computing continuum. While challenges such as resource constraints and scalability persist, edge-centric approaches offer a viable path to enabling efficient and adaptable GenAI systems for future applications.

\def\refname{References}

\bibliographystyle{IEEEtran}
\bibliography{reference}

\end{document}